\documentclass[twocolumn,showpacs,aps,pre]{revtex4} 

\usepackage{graphicx}%
\usepackage{dcolumn} 
\usepackage{amsmath} 
\usepackage{latexsym} 
 
\begin {document} 
 
\title 
{ 
The International Trade Network: weighted network analysis and modelling
} 
\author 
{ 
K. Bhattacharya$^{1}$, G. Mukherjee$^{1,2}$, J. Saram\"aki$^3$, K. Kaski$^3$, and S. S. Manna$^{1,3}$ 
} 
\affiliation 
{ 
$^1$Satyendra Nath Bose National Centre for Basic Sciences 
    Block-JD, Sector-III, Salt Lake, Kolkata-700098, India \\ 
$^2$Bidhan Chandra College, Asansol 713304, Dt. Burdwan, West Bengal, India \\ 
$^3$Laboratory of Computational Engineering, Helsinki University of Technology, P.O. Box 9203, 
FIN-02015 TKK, Finland 
} 
\begin{abstract} 
 
Tools of the theory of critical phenomena, namely the scaling analysis and universality, are 
argued to be applicable to large complex web-like network structures. Using a detailed 
analysis of the real data of the International Trade Network we argue that the scaled link 
weight distribution has an approximate log-normal distribution which remains robust over 
a period of 53 years. Another universal feature is observed in the power-law growth of the 
trade strength with gross domestic product, the exponent being similar for all countries. 
Using the 'rich-club' coefficient measure of the weighted networks it has been shown that the 
size of the rich-club controlling half of the world's trade is actually shrinking. 
While the gravity law is known to describe well the social interactions in the static 
networks of population migration, international trade, etc, here for the 
first time we studied a non-conservative dynamical model based on the gravity law 
which excellently reproduced many empirical features of the ITN.

\end{abstract} 
 
\maketitle 
 
   Are the large real or man-made networks with complicated and heterogeneous connections
scale invariant and obey the well known scaling analysis and universality concepts
of Statistical Physics? In the recent years, extensive research effort have been devoted to
analyzing the structure, function and dynamics of complex networks relevant to 
multi-disciplinary fields of science~\cite{Barabasi,Dorogov,social}. Indeed the scale-free
networks reflect such scale invariance in the link-node structures of the
electronic communication networks like the Internet, world-wide web, protein interaction
networks and even in research collaboration networks. More recently it has been
observed that strengths of links of the networks, as called weights in graph theory,
also have very interesing properties and can shed much light into the understanding
of the details of the network~\cite{BBVPNAS,Barrat1,Manna,WL,Havlin}. Lately it has been 
proposed that the International Trade Network (ITN) i.e., the system of mutual
trading between different countries in the world can also be viewed as 
an interesting example of real-world 
network~\cite{Serrano,Garlaschelli,Garlaschelli1,Serrano1,Park,JariClustering}.

   In this paper we study the ITN as an excellent example of the weighted networks 
obeying the scale invariance and universality properties
where the extent of trade between a pair of countries can be treated as the link 
weight~\cite{Data,Gleditsch}. It is observed that the suitably scaled link weight distribution over 
many years can be approximated well by a log-normal law. The nodal strength measuring 
the total volume of trade of a country is seen to depend non-linearly on the country's 
Gross Domestic Product with a robust exponent. 
Many features observed in the analysis can be explained with a simple dynamical model, 
which has the well-known gravity model of international trade as its starting 
point~\cite{Tinbergen}.  

   In the ITN, a node depicts 
a country and an undirected link exists between any pair of nodes if the trade volume 
between the corresponding countries is non-zero. Both the number of nodes $N$ and 
the number of links $L$ in the ITN's show annual variation and grow almost 
systematically over 53 years, e.g. in 1948 $N=76$ and $L=1494$ whereas in 
2000 $N=187$ and $L=10252$. On the other hand the link density 
$L/\left[N(N-1)/2\right]$ is observed to remain roughly constant with values 
around 0.52 for the same period. 
The annual trade data are expressed in millions of dollars ($M\$$) of imports 
and exports between countries $i$ and $j$ using four different quantities 
${\rm exp}_{ij}$, ${\rm exp}_{ji}$, ${\rm imp}_{ij}$ and ${\rm imp}_{ji}$~\cite{Data,Gleditsch}. 
Due to differences in the reporting procedures, there are usually small deviations 
between exports ${\rm exp}_{ij}$ from $i$ to $j$ and imports ${\rm imp}_{ji}$ 
to $j$ from $i$. Therefore, we define the link weight $w_{ij}$, as a measure of 
the total trade volume between the two countries (in $M\$$), as follows   
\begin {equation} 
w_{ij}= ({\rm exp}_{ij}+{\rm exp}_{ji}+{\rm imp}_{ij}+{\rm imp}_{ji})/2. 
\end {equation} 
This quantity tends to average out the aforementioned discrepancies. The distribution 
of weights is observed to be broad with the smallest non-zero trade volume being 
less than 1 $M\$$ and the largest of the order of $10^5 M\$$. The number of links 
with very small weights is quite large, whereas only a few links with very large 
weights exist (Fig.~1). The tail of the distribution consists of links connecting very 
few high-income countries~\cite{HDR}.
The average weight per link is observed to grow during the investigated 
period, from 15.54 $M\$$ in 1948 to 308.8 $M\$$ in 2000. 
\begin{figure}[top] 
\begin{center} 
\includegraphics[width=7.0cm]{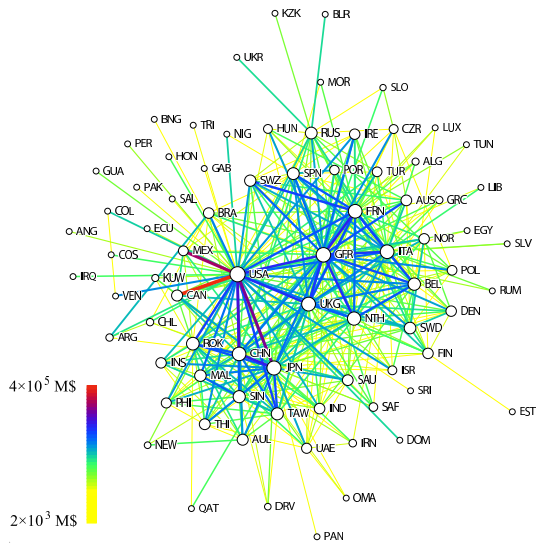} 
\end{center} 
\caption{(Color online) 
A subnetwork of the ITN for the year 2000, where only links with the highest 
4 $\%$ of weights and the associated nodes (countries) are included~\cite{Himmeli}, yielding in total 
80 nodes and 411 links. The node size is proportional to its strength and link color 
to its weight. Link weights are defined as the volume of annual trade between two 
countries in $M\$$. 
} 
\end{figure} 

   The probability density distribution of the link weights is defined as
the probability ${\rm Prob}(w)dw$ that a randomly selected link has 
the weight between $w$ and $w+dw$. The probability decays systematically 
with increasing weight and has a long tail with considerable fluctuation. We note 
that inferring the form of a broad probability distribution based on a relatively 
small number of data points has its difficulties; in the case of ITN, log-log plots 
of the distributions (not shown) display small intermediate linear regions which 
are (too) often interpreted as power laws. However, the tails of the distributions 
show clear curvature and much better fits to data are obtained by using a log-normal 
distribution of the form  
\begin {equation} 
{\rm Prob}(w)=\frac{1}{\sqrt{2\pi\sigma^2}}{1\over w}\exp \left(-\frac{\ln^2(w/w_0)}{2\sigma^2}\right), 
\end {equation} 
where the constants defined as $w_0 = \exp(\langle \ln(w) \rangle)$ and  
$\sigma = \{\langle (\ln(w))^2 \rangle - \langle \ln(w) \rangle ^2\}^{1/2}$ are observed 
to have different values for the weight distributions for different years. However, 
one can get a plot independent of $w_0$ and $\sigma$ by drawing 
$-2\sigma^2 \ln [{\rm Prob}\{\ln(w)\}\sqrt{2\pi\sigma^2}]$ as a function of $\ln(w/w_0)$, 
which gives a simple parabola $y=x^2$ for all years (Note that 
${\rm Prob}\{\ln(w)\}d\{\ln(w)\}={\rm Prob}(w)dw$ implies 
${\rm Prob}\{\ln(w)\}=w{\rm Prob}(w)$). 
In Fig.~2 we show such a plot, where the data has been aggregated for the five-year 
periods to reduce noise, i.e., 1951-55, 1956-60, ..., 1996-2000, each represented 
with its own symbol. It is clearly seen that the data points for each period fall close 
to the parabola $y=x^2$, displayed as the solid line. Therefore, we conclude that the 
annual weight distributions are reasonably well approximated by the log-normal 
distribution. This log-normal distribution of trade volumes has been discussed also in 
\cite {Fagiolo}. Note also that the trade imbalances have been reported to follow a 
log-normal distribution~\cite{Serrano1}.  
\begin{figure}[top] 
\begin{center} 
\includegraphics[width=7.5cm]{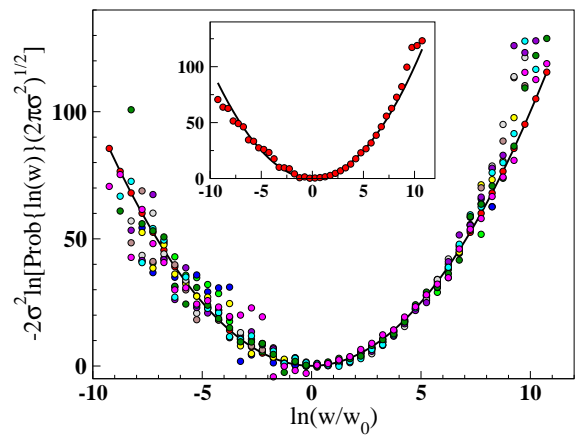} 
\end{center} 
\caption{(Color online) 
The probability distribution of link weights of the ITN. This plot displays 
$-2\sigma^2 \ln [{\rm Prob}\{\ln(w)\}\sqrt{2\pi \sigma^2}]$ as a 
function of $\ln(w/w_0)$. The data points scatter around the parabola $y=x^2$ 
(solid line), which implies a log-normal probability distribution (see text). 
Different symbols depict empirical distributions where the data have been  
aggregated over five-year periods. The inset shows the distribution of all link 
weights over the time span of 50 years from 1951 to 2000. 
} 
\end{figure} 

   For weighted complex networks, the strength $s_i$ of node $i$ is defined 
as $s_i = \Sigma_j w_{ij}$~\cite{BBVPNAS}, 
which in the case of ITN corresponds to the total volume of annual trade 
associated with a node. Intuitively, one can expect that in general the strengths 
of high-GDP. 
countries are higher than those of the low-GDP 
countries. To see this in a quantitative fashion, we have utilized an elastic 
constant $\gamma$ to measure how changes in strength respond to changes in GDP
(we have used total Gross Domestic Product of the country and not the GDP per capita
without taking into account of inflation).
Hence, we define $\gamma$ as:
$(ds/s)/(dG/G) = \gamma$, which results in a power-law relationship $s_i  \propto G^{\gamma}_i$. 
In Fig.~3 we plot the strengths $s_i$ vs. GDP $G_i$ for 22 different countries, 
representing a good mix of economic strengths, i.e., 10 high, 5 middle and 7 
low income countries. It is observed that the strength of each country grows 
non-linearly with its GDP with approximately the same slope. In the inset of this 
figure we show the probability distribution of $\gamma$ values for 168 different
countries. The distribution has a long tail and the $\gamma$ values of 12
countries are found to be larger than 2. A detailed inspection reveals that majority of these 12
countries are those originated after Soviet Union, Yugoslavia and Czechoslovakia were fragmented.
An overall average of the growth exponents has been estimated to be $\gamma = 1.26$
which comes down to 1.06 if these 12 countries are not considered for the averaging. 
The peak value of the distribution occurs very close to $\gamma=1$. Interestingly, 
Irwin~\cite{Irwin} has observed earlier that the total world export volume varies 
as a function of the total world real GDP to the power of 1.16, along with other 
factors. In comparison, however, our observation reveals a more detailed picture, 
indicating that the total trade volumes of individual countries are also 
approximately power laws as function of their individual GDPs, with exponents 
close to this value. 
\begin{figure}[top] 
\begin{center} 
\includegraphics[width=6.5cm]{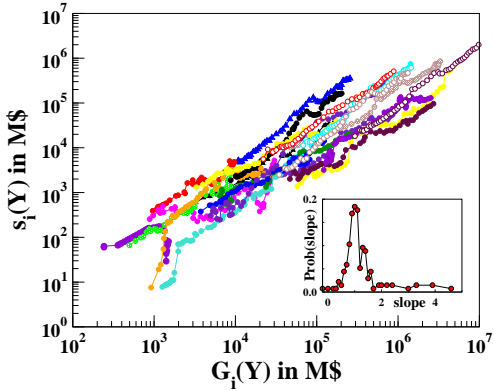} 
\end{center} 
\caption{(Color online) 
The dependence of strength $s_i(Y)$ of country $i$ measuring its total volume of trade
on its GDP $G_i(Y)$ in year $Y$, plotted for each country (color coded) over the 53 
year span from 1948 to 2000. General trend corresponds to a non-linear growth
with an average exponent of around $1.2$.
The inset shows the distribution of exponents associated with individual countries. 
} 
\end{figure} 

   Non-trivial correlations among the nodes of different real-world networks have
been observed. For an unweighted network this means that large degree nodes are 
connected among themselves forming a club. More precisely such a club consists of a
subset of $n_k$ nodes whose degree values are at least $k$. A rich-club coefficient (RCC)
is measured as $\phi(k) = 2E_k/[n_k(n_k-1)]$ where $E_k$ is the number of links 
actually exists in the club and $[n_k(n_k-1)]/2$ is the total number of node pairs
in the club \cite {Colizza}. A high value of $\phi(k)$ implies that members are indeed tightly connected. 
However it has been realized that only this definition is not enough,
since with this measure even uncorrelated random graphs show some rich-club
effect as well \cite {Colizza}. It is suggested that one needs to define a `null model' or the maximally
random network (MRN) keeping the nodal degree values $\{k_i\}$ preserved, measure the 
corresponding RCC $\phi_{ran}(k)$ and observe the variation of ratio $\rho(k)=\phi(k) / \phi_{ran}(k)$.
We have executed the same analysis for the ITN of the year 2000
and generated the MRN using the pairwise linkend exchange method \cite {Colizza}. However it is observed
that the variations of $\phi(k)$ and $\phi_{ran}(k)$ with $k$ are nearly same and consequently
$\rho(k)$ is nearly equal to unity for the whole range of degree values.

Next we studied the rich-club effect of the same ITN but now considering it as
a weighted network. The rich-club is now defined as the subset of nodes whose
strengths are at least $s$ controlling a major share of the world's trade 
dynamics. The RCC of the weighted network is defined as:
\begin {equation} 
R_w(s) = 2\Sigma_{i,j}w_{ij}/[n_s(n_s-1)], 
\end {equation} 
The corresponding maximally random weighted network (MRWN) has been 
generated keeping both the nodal degrees $\{k_i\}$ as well as the nodal strength 
values $\{s_i\}$ preserved. 
   
To generate the MRWN from the MRN of the ITN of the year 2000 described above
a self-consistent iteration procedure is used to obtain the link weight distribution
consistent with the nodal strength list $\{s_i\}$. We start assigning arbitrary random 
numbers as the weights $w_{ij}$ to all links maintaining that the weight matrix is always symmetric i.e.,
$w_{ij}=w_{ji}$. For an arbitrary node $i$ the difference $\delta_i=s_i-\Sigma_jw_{ij}$ is calculated.
Weights of all $k_i$ links meeting at the node $i$ are then updated as
\begin {equation}
w_{ij} \rightarrow w_{ij} + \delta_i \left(w_{ij}/\Sigma_jw_{ij}\right).
\end {equation}
to balance $s_i$ and $\Sigma_jw_{ij}$. By repeated iterations the link weights quickly
convergence and attain consistency with nodal strengths $\{s_i\}$. It is checked that
$\langle s_is_j \rangle \sim w_{ij}$ relation is well satisfied for this MRWN.

\begin{figure}[top] 
\begin{center} 
\includegraphics[width=6.0cm]{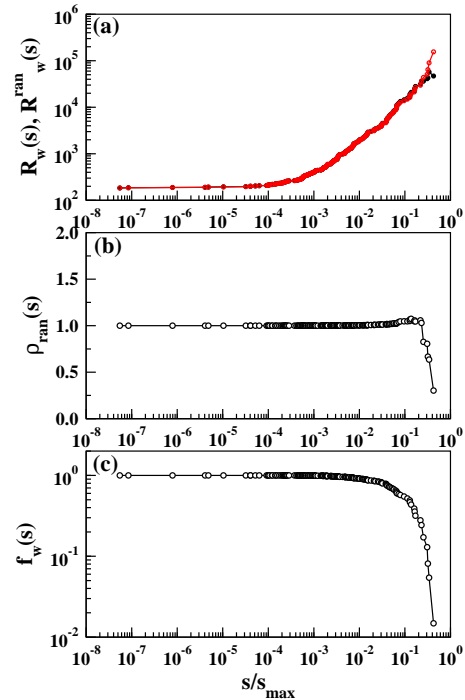} 
\end{center} 
\caption{(Color online) (a) The weighted rich-club coefficients $R_w(s)$ (black) and $R^{ran}_w(s)$ (red)
for the ITN of the year 2000, (b) The ratio $\rho_{ran}(s)$ of these two coefficients
is nearly equal to unity over the whole range of variation and (c)
$f_w(s)$, of the international trade volume among the rich-club 
members having strengths $s$ and above.
} 
\end{figure} 

In Fig. 4(a) we plot both $R_w(s)$ for the ITN and $R^{ran}_w(s)$ for the MRWN with
the scaled nodal strength $s/s_{max}$. The two measures are found to be nearly 
same, grow like $s^{0.85}$ for large $s$ values and their ratio $\rho_{ran}(s)$ is nearly equal to one 
except for a few values of $s$ near $s_{max}$ (Fig. 4(b)).

To explain this we observe that only 15$\%$ elements of the adjacency matrices of the
ITN and the MRN are different. Therefore a typical node of ITN retains the links to most
of its neighbors even after maximal randomization. This is because of the high value of the
link density ($59\%)$ of the ITN for the year 2000. As a result $\rho_{ran}(k)$ as well
as $\rho_{ran}(s)$ are nearly equal to unity. This implies that the pairwise link connections 
and the associated link weights of the original ITN are very close to those 
of the corresponding MRWN. In fact one can say that the original ITN
is a typical member of the different random configurations of the MRWN when the $\{k_i\}$ as well as
$\{s_i\}$ sets of the ITN are preserved. These results on the randomized unweighted ensemble confirms the
results in \cite {Garlaschelli} when compared with randomized networks with fixed degree
sequence as in \cite {Park}.

Zhou and Mondrag\'on \cite {Zhou} observed a very similar behavior for the Internet statistics. 
Following them we conclude that rich nodes in the original ITN and in its corresponding
MRN and MRWN are tightly connected and the similarity of the rich-club connectivity in the ITN structure
(with and without weights) does not imply that ITN lacks a rich-club structure.

In fact the presence of the rich-club effect is evident even if we simply analyze the variation of the
fraction $f_w(s)$ of the total volume of trade taking place 
among the members of the club. This is depicted in Fig.~4(c) which shows that $f_w(s)$ 
remains close to unity until high values of $s/s_{max}$, then decreases gradually 
to 1/2 at $s/s_{max} \approx 0.11$, and finally drops sharply. Very interestingly 
we also observe that the size of the rich club containing 50$\%$ of the total volume 
of annual trade shrinks almost systematically from 19$\%$ in 1948 to 8$\%$ in 2000. 

\begin{figure}[top] 
\begin{center} 
\includegraphics[width=6.5cm]{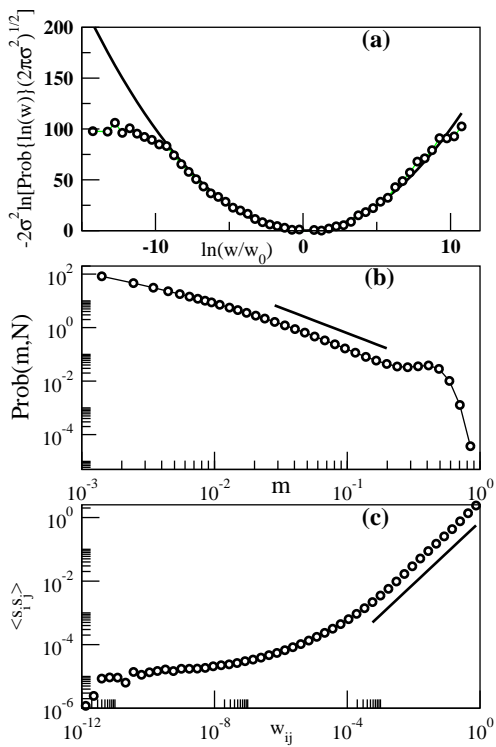} 
\end{center} 
\caption{Model results: (a) Scaled weight distribution fits well with the simple parabola 
$y=x^2$. (b) The GDPs of the individual countries have a broad distribution, which in the 
tail region seems to follow roughly a power law behavior (shown as a line with exponent $\sim$1.92). 
(c) The strength correlation $\langle s_is_j \rangle$ as a function of 
link weight is shown to grow approximately as $\langle s_is_j\rangle \propto w_{ij}^{0.98}$. 
} 
\end{figure} 
 
   The very heterogeneous distribution of trade volumes in the ITN is also reflected 
in the average pair correlation function $\langle s_is_j \rangle$ of nodal strengths 
and in its power law dependence on the link weights $w_{ij}$. Links with high weights 
$w_{ij} \sim w_{max}$ obviously must connect pairs of nodes of high strength, and for 
them $\langle s_is_j \rangle \sim s_{max}^2$. On the other hand for links of weights 
around unity, $\langle s_is_j \rangle \sim s_{max}$. Assuming that $w_{max}$ itself 
is of the order of $s_{max}$, we find an upper bound for the exponent $\nu=1$ 
describing the variation of $\langle s_is_j \rangle \sim w_{ij}^{\nu}$. Our analysis 
of the ITN yields, however, a somewhat smaller value of $\nu$ being between 0.65 and 
0.90 for different financial years between 1948 and 2000. The dependence of the weight 
distribution on the underlying topological structure of the ITN is studied by measuring 
the average strength of a node as a function of its degree, which turn out to exhibit 
strong degree of non-linearity: $\langle s(k) \rangle \propto k^{\mu}$ where $\mu$ 
varies between 3.4 and 3.7 for the same period. 

   We now develop a dynamical model based on the well-established gravity 
model~\cite {Tinbergen} used in social and economic 
sciences to describe the flow of social interaction between two economic centers 
$i$ and $j$ as a function of their economic sizes $m_i, m_j$ and distance $\ell_{ij}$:  
$F_{ij} = Gm_im_j/\ell_{ij}^2$. This equation has been generalized 
to the parametric form~\cite{Head} 
\begin{equation} 
F_{ij}={m_i^{\alpha}} \left({\frac{m_j^{\beta}}{\ell_{ij}^{\theta}}}/ 
{\Sigma_{k \ne i} \frac{m_k^{\beta}}{\ell_{ik}^{\theta}}}\right). \label{GravityEq} 
\end{equation} 
where the exponents $\alpha$ and $\beta$ usually range between 0.7 and 1.1 where as
$\theta$ is observed to be around 0.6 \cite {Head}.

   In our model, we assume a unit square to represent the world and N points distributed 
at random positions representing the capital cities of different countries. Initially 
the GDP values $m_i (i=1,N)$ are randomly assigned with uniform probability such that 
the total GDP is unity: $\Sigma_{i=1,N} m_i=1$. Then we let the dynamics start, which 
is essentially a series of pairwise interactions. There are different pairwise wealth 
exchange models studied in the literature which calculate the distribution of wealths
in a society \cite {Wealth}. A pair of countries $(i,j)$ is randomly 
selected $(1 \le i,j \le N)$ for a transaction and time $t$ is measured as the number 
of such transactions. In a transaction the selected countries invest the amounts 
$F_{ij}$ and $F_{ji}$ calculated using Eq.~(\ref{GravityEq}). Then the total amount 
of investment $\tilde F_{ij} = F_{ij}+F_{ji}$ is randomly shared between the two 
countries as a result of this transaction, as follows: 
\begin {eqnarray} 
   m_i=m_i-F_{ij}+\epsilon \tilde F_{ij}+\Delta_i, \\ 
   m_j=m_j-F_{ji}+(1-\epsilon)\tilde F_{ij}+\Delta_j. 
\end {eqnarray} 
Here $\epsilon$ is a random fraction freshly drawn for every transaction. 
The random sharing of $\tilde F_{ij}$ is justified by the fact that while
the Gravity law describes the average interaction in terms of the strengths and
distances of separation, the actual amount of trade depends on other factors,
many of them are political. 
In this 
idealized model countries are not allowed to make debt, which in turn makes the dynamics 
non-conservative through the parameters $\Delta_i$ and $\Delta_j$. It holds for these 
parameters that $\Delta_i=0$ if 
$F_{ij} < m_i$ and $\Delta_j=0$ if $F_{ji} < m_j$. However, if for some transaction  
$F_{ij} > m_i$ or $F_{ji} > m_j$ then we add $\Delta_i = F_{ij} - m_i$ or 
$\Delta_j = F_{ji} - m_j$ such that after the transaction, GDP balance does not 
become negative. Also after the transaction the individual GDP's are rescaled 
$m_i \to m_i/{\Sigma_j m_j}$ for the total GDP to remain unity. It is observed that 
a large number of pairwise transactions leads to a stationary state where $\langle m^2 \rangle$ 
fluctuates with time around a steady mean value. Any time after reaching the stationary state, the 
dynamics is used to construct a model ITN such that links are established between 
countries $i$ and $j$ whenever there is a transaction between them. We let the 
dynamics run until a pre-assigned link density (typically 0.3-0.5) has been reached. 
For example to generate a network corresponding to the ITN of the year 2000, we take
$N$ = 187 and continue the exchange dynamics till $L$ = 10252 distinct links are dropped
corresponding to the link density $0.59$.
The weight of a link is then defined as the total amount of investment between pairs 
of countries in all transactions. 
 
  For comparison the weight distribution of our model networks is analyzed in the 
same way as the real ITN data, and it turns out that an excellent consistency with 
the simple parabola $y=x^2$ is observed for parameter values 
$\alpha=1/2, \beta=1, \theta=1/2$, within a tolerance of 0.2 for all exponents, 
see Fig. 5. We also find that the 
GDP distribution is broad showing a power law like behavior for a short interval 
of $m$ (with exponent $\sim$1.9) before finite size effects set in. This is to be 
compared with the real-world data where the GDP distribution of different countries 
has been argued to be consistent with the Pareto law \cite{Pareto}. Finally, as in 
the real ITN, the two-point strength correlation is seen to grow 
like $\langle s_is_j \rangle \propto w_{ij}^{\nu}$  with $\nu \approx 0.98$ in the 
large weight limit, compared to the range of values 0.65 - 0.90 in the real ITN. 
In addition we find that the cumulative degree distribution is consistent with the 
results on the real ITN. 

   To summarize, we have shown that the weighted International Trade Network can be
looked upon as an excellent example of a complex network obeying scale-invariance
and universality features. The scaled distributions of annual world trade volumes 
between countries collapse well to a log-normal distribution and it remains
unchanged over a span of 53 years implying robustness or universality.
Secondly, the nodal strength measuring the total trade volume associated 
with a country grows non-linearly with its GDP with a robust 
exponent. Also a large fraction of the global trade is controlled by a club
of few rich countries which shrinks its size as time goes on.
Finally, the main features 
of the real-world ITN have been reproduced by using a simple non-conservative dynamical 
model starting from the well-known gravity model of social and economic sciences. 

   We thank K. S. Gleditsch, A. Chakrabarti for useful communications. SSM thanks LCE, HUT 
for a visit and hospitality. JS and KK acknowledge the support from the Academy of 
Finland's Centers of Excellence programs for 2000-2005 and 2006-2011. 

\leftline {Electronic Address: manna@boson.bose.res.in}

\end {document}